\documentclass[reprint,superscriptaddress,floatfix,prb,nolongbibliography]{revtex4-2}
\usepackage{graphicx}
\usepackage{textcomp}
\usepackage{siunitx}
\usepackage{bm}
\usepackage[normalem]{ulem}

\usepackage{amsmath}
\usepackage{amsfonts}
\usepackage{amssymb}
\usepackage{xcolor}
\usepackage[utf8]{inputenc}
\usepackage[hidelinks]{hyperref}

\newcommand{\black}[1]{\textcolor{black}{#1}}

\newcommand{\iu}{{i\mkern1mu}}

\newcommand{\mathperiod}{\,.} % end of sentence in mathematical environment
\newcommand{\mathcomma}{\,,}  % comma in mathematical environment

\newcommand{\Tc}{T_\mathrm{c}}
\newcommand{\norm}[1]{\left\lVert #1 \right\rVert_2}
\newcommand{\gsim}{\gtrsim}
\newcommand{\lsim}{\lesssim}
\newcommand{\RN}[1]{%
  \textup{\uppercase\expandafter{\romannumeral#1}}%
}
\begin{document}

\title{Bosonic excitation spectra of superconducting $\mathrm{Bi_2Sr_2CaCu_2O_{8+\delta}}$ and $\mathrm{YBa_2Cu_3O_{6+x}}$ extracted from scanning tunneling spectra}

\author{Thomas Gozlinski}
\affiliation{Physikalisches~Institut,~Karlsruhe~Institute~of~Technology,~76131 Karlsruhe,~Germany}
\author{Mirjam Henn}
\affiliation{Physikalisches~Institut,~Karlsruhe~Institute~of~Technology,~76131 Karlsruhe,~Germany}
\author{Thomas Wolf}
\affiliation{Institute~for~Quantum~Materials~and~Technologies,~Karlsruhe~Institute~of~Technology,~76344~Eggenstein-Leopoldshafen,~Germany}
\author{Matthieu Le Tacon}
\affiliation{Institute~for~Quantum~Materials~and~Technologies,~Karlsruhe~Institute~of~Technology,~76344~Eggenstein-Leopoldshafen,~Germany}
\author{Jörg Schmalian}
\affiliation{Institute~for~Quantum~Materials~and~Technologies,~Karlsruhe~Institute~of~Technology,~76344~Eggenstein-Leopoldshafen,~Germany}
\affiliation{Institute~for~Theory~of~Condensed~Matter,~Karlsruhe~Institute~of~Technology,~76131~Karlsruhe,~Germany}

%\affiliation{Institut~für~Festkörperphysik,~Karlsruhe~Institute~of~Technology,~76344~Eggenstein-Leopoldshafen,~Germany}
\author{Wulf Wulfhekel}
\affiliation{Physikalisches~Institut,~Karlsruhe~Institute~of~Technology,~76131 Karlsruhe,~Germany}
\affiliation{Institute~for~Quantum~Materials~and~Technologies,~Karlsruhe~Institute~of~Technology,~76344~Eggenstein-Leopoldshafen,~Germany}

\date{\today}

\begin{abstract}
A detailed interpretation of scanning tunneling spectra obtained on unconventional superconductors enables one to gain information on the pairing boson. Decisive for this approach are inelastic tunneling events. Due to the lack of momentum conservation in tunneling from or to the sharp tip, those are enhanced in the geometry of a scanning tunneling microscope compared to planar tunnel junctions. This work extends the method of obtaining the bosonic excitation spectrum by deconvolution from tunneling spectra to nodal $d$-wave superconductors. In particular, scanning tunneling spectra of slightly underdoped $\mathrm{Bi_2Sr_2CaCu_2O_{8+\delta}}$ with a $\Tc$ of \SI{82}{\kelvin} and optimally doped $\mathrm{YBa_2Cu_3O_{6+x}}$ with a $T_c$ of \SI{92}{\kelvin} reveal a resonance mode in their bosonic excitation spectrum at $\Omega_\mathrm{res} \approx \SI{63}{\milli e\volt}$ and $\Omega_\mathrm{res} \approx \SI{61}{\milli e\volt}$ respectively. In both cases, the overall shape of the bosonic excitation spectrum is indicative of predominant spin scattering with a resonant mode at $\Omega_\mathrm{res}<2\Delta$ and overdamped spin fluctuations for energies larger than $2\Delta$. To perform the deconvolution of the experimental data, we implemented an efficient iterative algorithm that significantly enhances the reliability of our analysis.
\end{abstract}

\maketitle

\section{Introduction}
With the intention to unravel the unconventional pairing mechanism in high-temperature superconductors, extensive effort has been put into extracting the spectral density of the pairing boson from experimental data \cite{zasadzinski_correlation_2001, zasadzinski_tunneling_2003, zasadzinski_persistence_2006, muschler_electronboson_2010, lee_interplay_2006, van_heumen_optical_2009, carbotte_bosons_2011, dal_conte_disentangling_2012, bok_quantitative_2016, chubukov_pairing_2020}. An ever-present contender for this ``bosonic glue'' are antiferromagnetic spin fluctuations which have been extensively studied in the family of cuprate superconductors \cite{rossat-mignod_neutron_1991, fong_neutron_1999, scalapino_case_1995, ogata_t_2008, scalapino_common_2012}. Such an electronic pairing mechanism leads to a heavy renormalization of the boson spectrum when entering the superconducting state. In the normal state, overdamped spin excitations form a broad and gapless continuum. In the superconducting state, they develop a spin gap of $2\Delta$, the minimum energy needed to create a particle-hole excitation, plus a rather long-lived resonance mode at $\Omega_\mathrm{res}<2\Delta$ inside the spin gap \cite{abanov_relation_1999, eschrig_neutron_2000, abanov_fingerprints_2001, eschrig_dispersion_2002, abanov_neutron_2002, eremin_novel_2005, eschrig_effect_2006, korshunov_theory_2008, maier_neutron_2009, yu_universal_2009, hlobil_strong_2013}. This resonance is made possible by the sign-changing ($d$-wave) symmetry of the superconducting gap and identified as a spin exciton \cite{abanov_fingerprints_2001}. The above mentioned behavior of the spin excitation spectrum has been directly observed in inelastic neutron scattering (INS) experiments \cite{rossat-mignod_neutron_1991, mook_polarized_1993, fong_phonon_1995, bourges_inelastic-neutron-scattering_1996, fong_neutron_1999, he_resonant_2001, he_magnetic_2002} yielding strong evidence for spin-fluctuation mediated pairing. Since signatures of this resonance mode are also expected to be visible in optical, photoemission and tunneling spectra, a considerable number of studies tried to complete the picture using these techniques \cite{hwang_evolution_2007, van_heumen_optical_2009, muschler_electronboson_2010, dal_conte_disentangling_2012, schachinger_finite_2008, carbotte_bosons_2011, bok_quantitative_2016, yamaji_hidden_2021, zasadzinski_tunneling_2003, lee_interplay_2006, zasadzinski_persistence_2006, sui_eliashberg_2015}, all probing a slightly different boson spectrum and facing complicated inversion techniques. Recently, machine learning algorithms entered the scene and their application to angle-resolved photoemission (ARPES) data proved to be a powerful concept to reverse-model the spin-spectrum, but this happens at the cost of a number of free parameters which cannot be easily mapped onto physical quantities \cite{chubukov_pairing_2020, yamaji_hidden_2021}. In this work, we extracted the bosonic spectrum from the inelastic part of scanning tunneling spectra which we obtained on the cuprate superconductors $\mathrm{Bi_2Sr_2CaCu_2O_{8+\delta}}$ (Bi2212) and $\mathrm{YBa_2Cu_3O_{6+x}}$ (Y123). In contrast to previous scanning tunneling spectroscopy (STS) \cite{zasadzinski_persistence_2006, sui_eliashberg_2015} and break junction experiments \cite{ahmadi_eliashberg_2011} that focused on Bi2212, we obtain the boson spectrum without a functional prescription and over a wide energy range for both materials. It naturally exhibits the sharp resonance mode and overdamped continuum that are characteristic for the spin spectrum measured in INS.

%Tunneling spectroscopy is a versatile tool to analyze the electronic structure of solids, especially when performed with lateral resolution in a scanning tunneling microscope (STM). Besides elastic tunneling of electrons, also inelastic tunneling may occur, in which the the total energy and momentum is shared in the final state between the inelastically scattered electron and a bosonic excitation.

Inelastic electron tunneling spectroscopy experiments using the tip of a scanning tunneling microscope (IETS-STM) have proven to be a powerful tool in the study of bosonic excitations of vibrational \cite{vitali_phonon_2004,lee_interplay_2006, fransson_surface_2007, gawronski_imaging_2008, schackert_local_2015}, magnetic \cite{li_kondo_1998, madhavan_tunneling_1998, stipe_single-molecule_1998, heinrich_single-atom_2004, balashov_magnon_2006, spinelli_imaging_2014} or plasmonic \cite{vitali_phonon_2004, chong_narrow-line_2016} character in metals, single molecules and also superconductors. Due to the spatial confinement of the electrons in the apex of the STM tip, the wave vector of the tunneling electrons is widely spread and the local density of states (LDOS) of the tip becomes rather flat and featureless. Consequently, the generally momentum dependent tunnel matrix element can be considered momentum independent in the STM geometry \cite{berthod_tunneling_2011}. As a result, the elastic contribution to the tunneling conductance $\sigma^\mathrm{el}$ becomes directly proportional to the LDOS of the sample, as has been shown by Tersoff and Hamann \cite{tersoff_theory_1985}. Similarly, the inelastic contribution $\sigma^\mathrm{inel}$ to the tunneling conductance is given by a momentum integrated scattering probability of tunneling electrons sharing their initial state energy with a final state electron and a bosonic excitation. The absence of strict momentum conservation opens the phase space for the excited boson and as a consequence, the inelastic contributions to the tunneling current can be a magnitude larger than in planar junctions \cite{schackert_local_2015}, in which the lateral momentum is conserved.

Previous IETS-STM experiments used this effect to determine the Eliashberg function $\alpha^2F(\Omega)$ of the strong-coupling conventional superconductor Pb \cite{schackert_local_2015,jandke_coupling_2016}, which contains the momentum integrated spectral density of the pairing phonon $F(\Omega)$, as well as the electron-phonon coupling (EPC) constant $\alpha(\Omega)$ \cite{mcmillan_lead_1965, eliashberg_interactions_1960, eliashberg_temperature_1961}. While for conventional superconductors, the Migdal theorem \cite{migdal_interaction_1958} allows to treat the electronic and phononic degrees of freedom to lowest order separately, largely simplifying the analysis of IETS spectra, the situation is less clear for unconventional superconductors with electronic pairing mechanism. Nevertheless, the theoretical description of IETS-STM spectra could be extended to the fully gapped Fe-based unconventional superconductors of $s^\pm$ character \cite{hlobil_tracing_2017,jandke_unconventional_2019}. 
%In these materials, the bosonic spectrum undergoes heavy renormalization when entering the superconducting state. In the normal state overdamped spin excitations form a broad and gapless continuum. In the superconducting state they become gapped by the full superconducting gap $2\Delta$ with exception of a rather long-lived resonance mode at around $1.3\Delta$ \cite{yu_universal_2009}. 
Strong coupling of electrons and spin fluctuations manifests in IETS-STM spectra as a characteristically lower differential conductance in the superconducting state compared to the normal state for energies slightly below $3\Delta$ \cite{hlobil_tracing_2017}. Also for these systems, the boson spectrum could be reconstructed from IETS-STM spectra by deconvolution \cite{jandke_unconventional_2019}. In this work, we investigate, in how far this method can be extended to nodal $d$-wave superconductors. To do this, we follow the path of a deconvolution of scanning tunneling data, using \textit{a priori} band structure for the normal state model and the inelastic scanning tunneling theory of unconventional superconductors derived by Hlobil et al. \cite{hlobil_tracing_2017}. 

%From a series of inelastic neutron scattering (INS) experiments on various cuprate superconductors, the universal relationship between magnetic resonance mode and superconducting gap of $\Omega_\mathrm{res}/{\Delta}=\SI{1.28(8)}{}$ was found \cite{yu_universal_2009}. 
%While the resonance mode is directly obtained in INS, more model assumptions, like the applicability of the spin-fermion model, have to be put in, to deduce the spin excitation spectrum from a tunneling spectrum. In addition to that, the momentum information is lost in tunneling experiments. 
%In previous attempts to model the Eliashberg function, \textit{a priori} knowledge on its shape had to be put in \cite{sui_eliashberg_2015}, which biases the result due to the ill-posedness of the deconvolution problem. Recently, the application of machine learning algorithms on angle-resolved photoemission (ARPES) data proved to be a powerful concept to reverse-model the spin-spectrum, but this happens at the cost of a number of free parameters which cannot be easily mapped onto physical quantities \cite{chubukov_pairing_2020, yamaji_hidden_2021}. 
%Instead, we follow the path of a deconvolution of scanning tunneling data, using \textit{a priori} band structure for the normal state model and the inelastic scanning tunneling theory of unconventional superconductors derived by Hlobil et al. \cite{hlobil_tracing_2017}.  
%The method of deconvolution is applied for cuprate superconductors with the additional complication of a nodal gap function.

\section{Outline of the extraction procedure}\label{sec:outline of the procedure}
The total tunneling conductance $\sigma^\mathrm{tot}$ between a normal conducting tip and a superconductor is comprised of the elastic tunneling contributions $\sigma^\mathrm{el}$, but also significant inelastic contributions $\sigma^\mathrm{inel}$ \cite{schackert_local_2015, jandke_coupling_2016,hlobil_tracing_2017, jandke_unconventional_2019}. While the second derivative of the tunneling current $\mathrm{d}^2I/\mathrm{d}U^2$ obtained on conventional superconductors in the normal state is directly proportional to the Eliashberg function $\alpha^2F(\Omega)$ \cite{jandke_coupling_2016}, the bosonic glue in unconventional superconductors is drastically renormalized upon entering the superconducting phase.

In the presence of strong inelastic contributions to the tunneling current an inversion procedure \textit{à la} McMillan and Rowell \cite{mcmillan_lead_1965} cannot be used to extract the Eliashberg function from the superconducting spectrum. As was shown in Ref. \cite{hlobil_tracing_2017}, the function $g^2\chi^{\prime\prime}(\Omega)$ acts as the ``generalized glue function'' and analog to the Eliashberg function in superconductors driven by electronic interactions. As both, phonons and spin fluctuations, may couple to the tunneling quasiparticles, we define the bosonic spectrum
\begin{equation}
    B(\Omega) \approx \alpha^2F(\Omega) + g^2\chi^{\prime\prime}(\Omega)
\end{equation}
where $g$ is the spin-fermion coupling constant and $\chi^{\prime\prime}$ is the dimensionless, momentum integrated spin spectrum. The inelastic differential conductance for positive voltage at zero temperature is given by
\begin{align}
    \sigma^\mathrm{inel}(eU) &\propto \int_0^\mathrm{eU} \mathrm{d}\Omega\, \nu_s(eU-\Omega) B(\Omega) \nonumber \\
    &=\left(\left[\nu_s\cdot \Theta\right]*\left[B\cdot\Theta\right]\right) (eU) \label{eq:convolution}.
\end{align}
 While the explicit momentum dependence of the bosonic spectrum is lost in this form, it still contains the spin resonance at the antiferromagnetic ordering vector if antinodal points on the Fermi surface contribute significantly to the tunneling spectrum. As can be seen from Eq.~\eqref{eq:convolution}, $B$ is the source function, the DOS in the superconducting state $\nu_s$ is the kernel and the inelastic tunneling conductance $\sigma^\mathrm{inel}$ is the signal. $\Theta$ denotes the Heaviside step function. The general aim in this work is to extract the function $B(\Omega)$ as accurately as we can from scanning tunneling spectra, which we do by deconvolution of Eq.~\eqref{eq:convolution}. We follow the following step-by-step procedure:
 \begin{enumerate}
     \item Determination of the superconducting density of states $\nu_s$
     \item Determination of the inelastic tunneling conductance $\sigma^\mathrm{inel}$
     \item Extraction of $B(\Omega)$ by deconvolution of Eq.~\eqref{eq:convolution}
 \end{enumerate}
 Assumptions and limitations of our extraction procedure are discussed in Section~\ref{subse:outline/assumptions and limitations}.

 \subsection{Step 1: Determining $\nu_s$}
 From a scanning tunneling spectrum below $T_c$ we obtain the differential conductance d$I/$d$U(eU)\equiv \sigma^\mathrm{tot}(eU)$. This function consists of the purely elastic part $\sigma^\mathrm{el}$ and the inelastic part $\sigma^\mathrm{inel}$. The elastic part is directly proportional to the superconducting density of states in the sample $\nu_s$. In this step, we determine the functional form of the elastic contribution by fitting a model function to the low bias region of the d$I/$d$U$ spectrum that keeps the complexity as low as possible while still capturing the relevant features of the band structure and pairing strength. We opt for a generalized Dynes function \cite{dynes_direct_1978} with a momentum dependent gap:
\begin{align}
    \sigma^\mathrm{el}_s(\omega)&=\frac{\sigma_0}{\mathcal{N}}\sum\limits_\lambda C_\lambda\int^{\pi/2}_0 \mathrm{d}\varphi  \nu^\mathrm{F}_\mathrm{n,\lambda}(\varphi) \times \nonumber\\
    &\times \left|\Re\left(\frac{\omega+\iu \Gamma_\lambda(\varphi)}{\sqrt{(\omega + \iu \Gamma_\lambda(\varphi))^2-\Delta_\lambda(\varphi)^2}}\right)\right|
    \mathperiod \label{eq:sc_dos_el}
\end{align}
Here, $\mathcal{N}=\sum_\lambda\int \mathrm{d}\varphi \nu^\mathrm{F}_\mathrm{n,\lambda}(k_F,\varphi)$ is a normalization factor, $\Delta_\lambda(\varphi)=\Delta_{0,\lambda}\cos(2\varphi)$ is the $d$-wave pairing potential, $\Gamma_\lambda(\varphi) = \gamma_\lambda\cdot |\Delta_\lambda(\varphi)|$ the quasiparticle scattering rate, $\nu^\mathrm{F}_\mathrm{n}(\varphi)$ is the angle-dependent DOS at the Fermi energy in the normal conducting phase, $\lambda$ the band index and $C_\lambda$ the relative tunneling sensitivity for the band. The function $\nu^\mathrm{F}_\mathrm{n}(\varphi)$ weighs gap distributions for different $(k_x,k_y)$ by their abundance along the Fermi surface (FS). It is derived from a microscopic tight-binding approach that models the dispersion relation. In the case of Bi2212 we used a single-band model whereas for Y123 we considered two CuO$_2$ plane bands and one CuO chain band (see Appendix~\ref{sec:app/calculated normal dos}). It should be noted that, unlike in fully gapped superconductors, the inelastic spectrum can be non-zero down to vanishing bias voltage because $\Delta(\bm{k})$ has a nodal structure. This prevents us from directly assigning the differential conductance for $e|U|\lsim \Delta$ to the purely elastic tunneling contribution as was possible for the $s_\pm$ superconductor monolayer FeSe \cite{jandke_unconventional_2019, hlobil_tracing_2017}. We will, however, start from here and refine $\sigma^\mathrm{el}$ in the next step.

\subsection{Step 2: Determining $\sigma^\mathrm{inel}$}\label{subsec:outline/Determining inelastic part}
 We use the fact that $\sigma^\mathrm{tot}(\omega)=\sigma^\mathrm{el}(\omega)+\sigma^\mathrm{inel}(\omega)$ and the physical constraints $\sigma^\mathrm{inel} (0) = 0$ and $\sigma^\mathrm{inel}(e|U|>0)>0$. For a slowly varying bosonic function, which we expect in the range $0<e|U|<\Delta$ due to the quick reopening of the gap near the nodal parts of the Fermi surface, $\sigma^\mathrm{inel}$ is essentially given by the elastic contribution $\nu_s(\omega)$ times some scalar, real factor. We thus assume, that, in the range $0<e|U|<\Delta$, $\sigma^\mathrm{el}$ is well guessed by our Dynes fit times a factor $\eta < 1$. We approximated $\eta$ using a boundary condition for the total number of states (see Appendix~\ref{subsec: app/The numerical scaling factor}) and in order to keep the condition $\sigma^\mathrm{el}(\infty) = \sigma_0$, we scale up our experimental curve by $1/\eta$ instead of scaling down our fitted curve. We took care that the choice of this numerical factor, that simply helps to perform our deconvolution algorithm and paint a more realistic picture of $\sigma^\mathrm{el}$, does not influence the extracted boson spectrum in a qualitative manner (see Appendix~\ref{subsec: app/The numerical scaling factor}).

\subsection{Step 3: Extracting $B(\Omega)$}
We compare two methods by which the boson spectrum was determined: The direct deconvolution in Fourier space and the iterative Gold algorithm \cite{gold_anl_1964, morhac_efficient_1997} to perform the deconvolution. The advantage of the Gold algorithm is that for positive kernel and signal, the result of this deconvolution method is always positive. This is in agreement with our physical constraint that the bosonic excitation spectrum is strictly positive. We used the implementation of the one-fold Gold algorithm in the TSpectrum class of ROOT system \cite{brun_root_1996, morhac_deconvolution_2006} in C language wrapped in a small python module.

The bosonic function from direct deconvolution in Fourier space is obtained from
\begin{equation}
    B(\Omega > 0) = \mathcal{F}^{-1}\left(\frac{\mathcal{F}(\sigma^\mathrm{inel})(t)}{\sqrt{2\pi}\mathcal{F}(\nu_s\cdot \Theta)(t)}\right)
\end{equation}
where $\mathcal{F}$ denotes a Fourier Transform and $\mathcal{F}^{-1}$ the inverse transform.
The abrupt change in elastic conductance at zero energy (multiplication with Heaviside distribution in Eq.~\eqref{eq:convolution}) leads to heavy oscillations in the Fourier components. Therefore the result of this deconvolution procedure can contain non-zero contributions for $E<0$ and negative contributions for $0<E<\Delta$. They are exact solutions to Eq.~\eqref{eq:convolution} but from a subset of non-physical solutions that we are not interested in. Because the solutions obtained in this way are highly oscillatory we show the result after Gaussian smoothing. In order to obtain a positive solution to Eq.~\eqref{eq:convolution}, the result of the direct deconvolution method is used as a first guess to the Gold algorithm. The results shown in this work are obtained after 2,000,000 iterations at which point convergence has been reached.
 
\subsection{Assumptions and limitations}\label{subse:outline/assumptions and limitations}
In order for our extraction method to be applicable, several simplifying assumptions were made:
\begin{enumerate}
    \item Quasiparticles with energy $\omega$ couple to bosonic excitations of energy $\Omega$ and effective integrated density of states $B(\Omega)$. The $k$-dependence of the interaction is thus disregarded.
    \item $\sigma^\mathrm{inel}$ has a simple relation to $\sigma^\mathrm{el}$ for $0<e|U|<\Delta$ (see Section~\ref{subsec:outline/Determining inelastic part}) which is generally oversimplified for d-wave superconductors, especially near $\Delta$.
\end{enumerate}
In general, retrieving the source function from a convolution integral is an ill-posed problem which means that we only obtain one solution from a large set of valid solutions to the convolution equation. 
Additional aspects that complicate the problem are the following:
\begin{enumerate}
    \item The kernel function $\nu_s(\omega)$ is a guess which is very dependent on the modeling of the superconducting density of states and is further questioned by lack of energetic regions of purely elastic processes in the scanning tunneling spectrum. We are in fact on the verge of a necessity for blind deconvolution algorithms.
    \item Strong electron-boson coupling leads to spectral features of $\sigma^\mathrm{el}$ outside the gap that are neglected here as the contribution of $\sigma^\mathrm{inel}$ is expected to be much larger. They can in principle be reconstructed within an Eliashberg theory using the extracted boson spectrum. Using this refined $\sigma^\mathrm{el}$ and repeating the procedure until $B(\Omega)$ leads to the correct $\sigma^\mathrm{el}$ and $\sigma^\mathrm{inel}$ could further improve our result.
    %\item Every involved function is non-zero for $U\rightarrow \infty$. However, for an efficient inversion of the problem, the fast Fourier transform method is unavoidable. Hence, all function arrays must be reasonably padded such that circular and linear convolution become similar. 
    \item Electronic noise in the recorded spectra is ignored and consequently ends up in either $\nu_s$ or $B(\Omega)$
    \item With $B(\Omega)$ we obtain only an ``effective tunnel Eliashberg function'' which includes all bosonic excitations that are accessible to the tunneling quasiparticles. Hence, no disentanglement into lattice and spin degrees of freedom is possible.
    \item The $k$-dependence of $B(\Omega)$ is inaccessible
\end{enumerate}

\section{Experimental methods}
We performed inelastic tunneling spectroscopy on a slightly underdoped Bi2212 sample with a $T_c$ of \SI{82}{\kelvin} (UD82) and an optimally doped Y123 sample with a $T_c$ of \SI{92}{\kelvin} (OP92) using a home-built STM with Joule-Thomson cooling \cite{zhang_compact_2011}. The samples were cleaved at a temperature of \SI{78}{\kelvin} in ultra-high vacuum (UHV) and immediately transferred into the STM. All spectra were recorded with a tungsten tip. The set-up also allows to vary the temperature of the STM in order to record spectra above $T_c$. Due to the large inhomogeneity of scanning tunneling spectra on Bi2212 \cite{fischer_scanning_2007}, we show averaged spectra recorded at positions, where the dip-hump feature can be clearly seen. In the case of Y123, the overall spectral inhomogeneity was lower (see Appendix~\ref{subsec: app/ldos inhomo}) and we show spectra which are averaged over a 50$\times$\SI{50}{\nano\metre^2} area where the dip-hump feature was ubiquitous.

\section{The case for $\mathrm{Bi_2Sr_2CaCu_2O_{8+\delta}}$}
\subsection{Experimental results}
\begin{figure}
    \centering
    \includegraphics{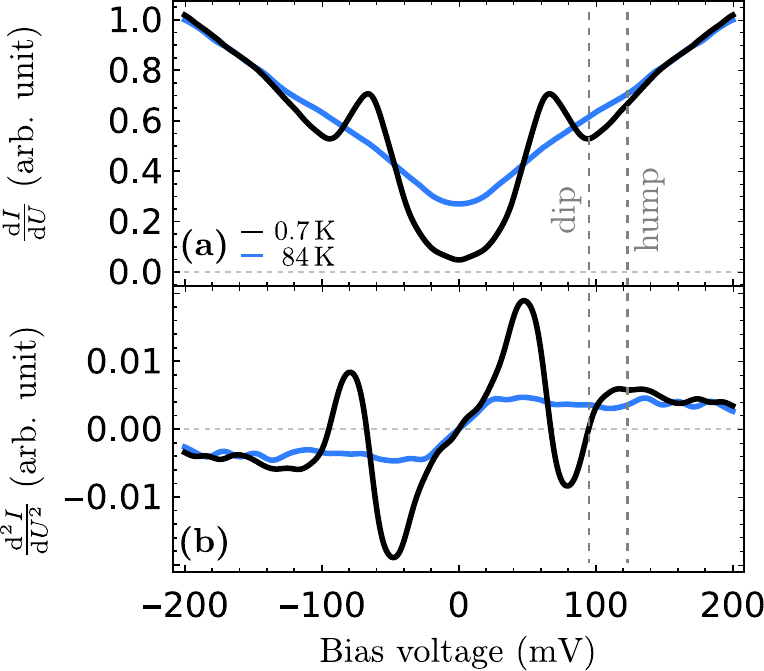}
    \caption{\textbf{Tunneling Spectra on Bi2212}: (a) Experimental $\mathrm{d}I/\mathrm{d}U$ spectra in the superconducting/normal state (black/blue) recorded at \SI{0.7}{}/\SI{84}{\kelvin} after Gaussian smoothing, symmetrization and normalization to the differential conductance in the normal state at \SI{200}{\milli\volt}. (b) Numerical derivative of spectra in (a).}
    \label{fig:dIdVnormsym}
\end{figure}
Fig.~\ref{fig:dIdVnormsym} shows experimental $\mathrm{d}I/\mathrm{d}U$ and $\mathrm{d}^2I/\mathrm{d}U^2$ spectra recorded at \SI{0.7}{\kelvin} and \SI{84}{\kelvin}. In order to remove a tilt in the spectra that stems from a slope in the density of states (DOS) due to hole doping \cite{randeria_particle-hole_2005, anderson_theory_2006}, we followed the standard procedure and symmetrized the spectra in Fig.~\ref{fig:dIdVnormsym}.
The d$I$/d$U$ spectrum for superconducting Bi2212 in Fig.~\ref{fig:dIdVnormsym}(a) shows a single but smeared gap with residual zero-bias conductance due to the nodal $d_{x^2-y^2}$ gap symmetry. Similarly, the coherence peaks are smeared due to the gap symmetry and possibly also due to short quasiparticle lifetimes. This is typical for the underdoped regime and may be caused by its proximity to the insulating phase \cite{fischer_scanning_2007}. Outside the gap, a clear dip of the superconducting spectrum below the normal conducting spectrum, followed by a hump reapproaching it, are visible. The V-shaped conductance in the normal state hints towards strong inelastic contributions to the tunneling current from overdamped electronic excitations that become partly gapped in the superconducting state as discussed in Ref. \cite{kirtley_inelastic-tunneling_1990, hlobil_tracing_2017}. The hump shows as a peak in the second derivative of the tunneling current that exceeds the curve of the normal state at $\approx \SI{120}{\milli \volt}$ in Fig.~\ref{fig:dIdVnormsym}(b). The relatively round shape of the superconducting gap in Bi2212 is atypical for a classic $d$-wave superconductor, in which the naive expectation is a V-shaped conductance minimum. As will be shown later on, the round shape of the gap can be generated without admixture of an $s$-wave pairing term by respecting the anisotropy of the Fermi surface in the normal state. The Fermi surface and gap anisotropy are summarized schematically in Fig.~\ref{fig:Fitting}(a).

\subsection{Extraction of the bosonic spectrum}
\begin{figure*}
    \centering
    \includegraphics{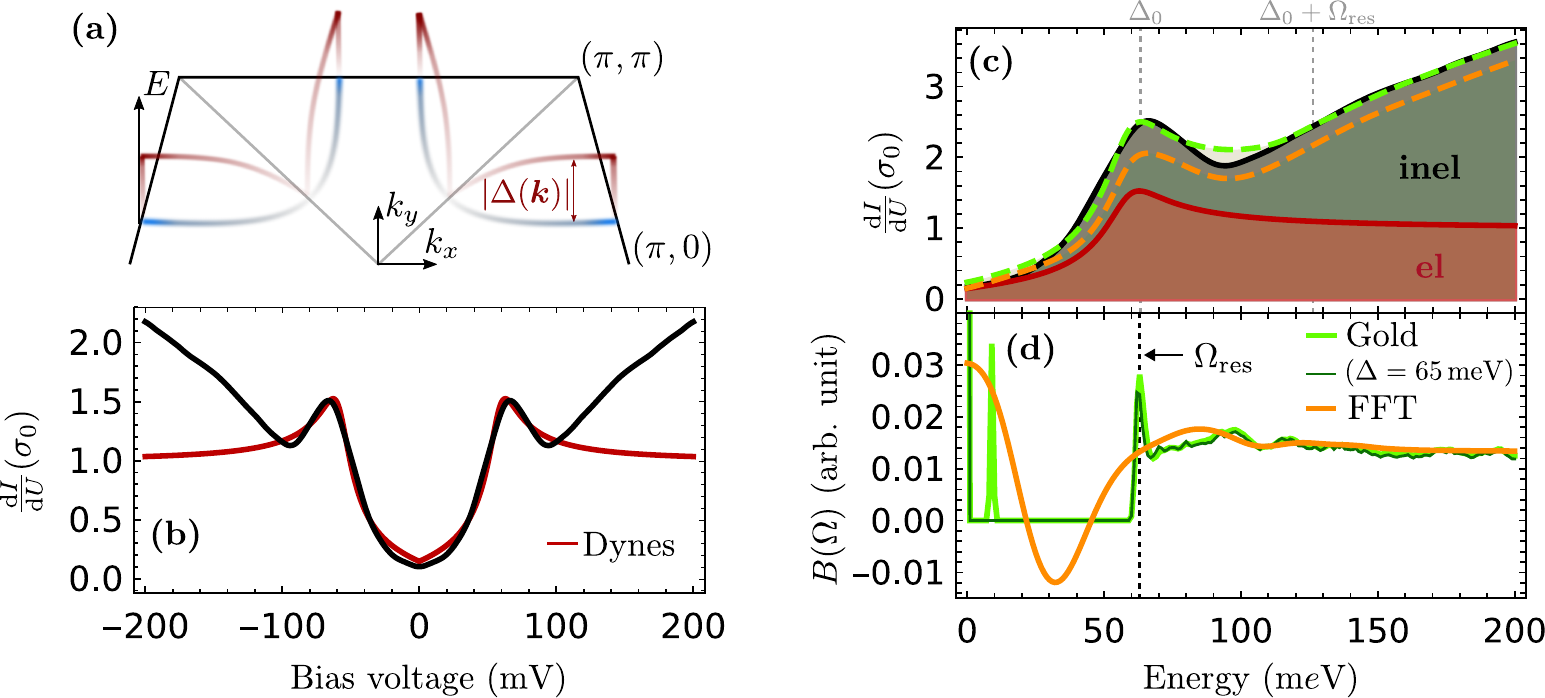}
    \caption{\textbf{Bosonic Spectrum Extraction for Bi2212}: (a) Schematic perspective view of the Fermi surface (FS) in the first Brillouin zone (BZ) in the normal/superconducting state (blue/red) (adapted from Ref. \cite{hoffman_imaging_2002}). The color lightness depicts the relative density of states: the higher the lightness, the lower the density of states. (b) A generalized Dynes model (Eq.~(\ref{eq:sc_dos_el})) with $\Delta_0 = \SI{63.30}{\milli e\volt}$ and $\gamma = \SI{0.15}{}$ (red) was fitted to the experimental differential conductance (black). (c) The total conductance (black line) has been scaled up by the factor $\eta^{-1}=1.67$. The inelastic part of the conductance (vertical width of grey area) is given by the difference between the total (black) and the elastic part (red) of the conductance. Forward convolved conductance with the obtained boson spectral functions from the direct FFT method/Gold algorithm are shown in orange/green dashed lines. (d) Boson spectral function determined by direct FFT method/Gold algorithm (orange/green). The thin dark green line shows the boson spectral function for a different Dynes fit than in (b,c) with $\Delta_0=\SI{65}{\milli e\volt}$ (not least square minimum). The result of the direct FFT method has been regularized for clarity. A clear resonance mode at $\Omega_\mathrm{res}\approx \SI{63}{\milli e \volt}$ is visible. Zero-energy contributions are an artefact from the scaling procedure.}
    \label{fig:Fitting}
\end{figure*}
We followed the step-by-step extraction procedure outlined in Section~\ref{sec:outline of the procedure} starting from the determination of the superconducting density of states $\nu_s$. The optimal Dynes fit to our experimental spectrum in the superconducting state is shown in Fig.~\ref{fig:Fitting}(b) with $\Delta_0 = \SI{63.31}{\milli e\volt}$ ($\Delta_\mathrm{max} = \SI{59.14}{\milli e\volt}$ and $\bar{\Delta}=\SI{49.60}{\milli e\volt}$) and $\gamma = \SI{0.19}{}$. The resulting (in)elastic contribution is shown in red(grey) in Fig.~\ref{fig:Fitting}(c). Here, a numerical scaling factor of $\eta = 0.6$ was used. The value for $\Delta$ lies within the range of previously reported gap values on the Bi2212 surface \cite{fischer_scanning_2007}, especially in the slightly underdoped regime, where variations of the local gap from the average gap tend to be larger \cite{mcelroy_coincidence_2005}.

The regularized bosonic function from direct deconvolution in Fourier space is shown in Fig.~\ref{fig:Fitting}(d) in orange. The contributions at low energies are an artefact from the scaling with factor $\eta$. Despite our uncompromising simplifications, the bosonic spectrum recovers well the tendency of the total conductance in the forward convolution (Fig.~\ref{fig:Fitting}(c) orange) and shows the expected behaviour for coupling to spin degrees of freedom at medium and high energies, i.e. a resonance mode at $\Delta < E < 2\Delta$ and approach of the normal state bosonic function $B$ for $E \gsim 3\Delta$ \cite{hlobil_tracing_2017}, that, in contrast to the Eliashberg function in the case of phonon-mediated pairing, remains finite for energies well above $2\Delta$. \black{While the long-lived resonance mode is associated with a spin resonance due to the sign-changing gap function, the broad high-energy tail of the bosonic spectrum is due to the coupling to overdamped spin fluctuations, or paramagnons \cite{abanov_fingerprints_2001}. Resonant inelastic x-ray scattering (RIXS) studies have shown that these paramagnons dominate the bosonic spectrum for energies larger than $\approx \SI{100}{\milli e\volt}$ in several families of cuprates as almost all other contributors, e.g. phonons, lie lower in energy \cite{le_tacon_intense_2011, minola_collective_2015, peng_dispersion_2018, wang_paramagnons_2022}. }

By application of the Gold algorithm we obtained the bosonic spectrum shown in green in Fig.~\ref{fig:Fitting}(d). Again, the high value at $E=0$ is a consequence of the scaling with factor $\eta$. The sharp peak at $\SI{10}{\milli e\volt}$ is due to inadequacies of our elastic fit in the region of the coherence peak. It is e.g. not present in our analysis of Y123 (see Section~\ref{sec:YBCO}) and vanishes once one takes an elastic DOS with a larger gap (here $\SI{65}{\milli e\volt}$, not least-square minimum) as we show in the thin dark green line in Fig.~\ref{fig:Fitting}(d). We could get rid of negative contributions and find a bosonic function that recovers well the total conductance (Fig.~\ref{fig:Fitting}(c) green), especially the dip-hump structure, and shows a very clear resonance at $\Omega_\mathrm{res} \approx \SI{63}{\milli e\volt}\approx 1.0 \Delta_0 \approx 1.1 \Delta_\mathrm{max}\approx 1.3 \bar{\Delta}$. 

The resonance mode extracted in this work is higher in energy than reported in inelastic neutron scattering experiments ($\Omega_\mathrm{res} \approx \SI{43}{\milli e \volt}$ at the antiferromagnetic ordering vector) \cite{fong_neutron_1999} and closer to the resonance determined by optical scattering ($\Omega_\mathrm{res} \approx \SI{60}{\milli e\volt}$) \cite{hwang_evolution_2007}. Due to the loss of momentum information in tunneling, the centre of the resonance is expected to be shifted to higher energies compared to the INS results \cite{hlobil_tracing_2017}. 
Due to the large inhomogeneity of $\Delta$ on the surface of Bi2212 \cite{fischer_scanning_2007}, it is more instructive to compare the ratio $\Omega_\mathrm{res}/\Delta$ to other works rather than the absolute value of $\Omega_\mathrm{res}$. The ratio $\Omega_\mathrm{res}/\bar{\Delta}=1.3$ lies within the current range of error of $\Omega_\mathrm{res}/\Delta=1.28\pm 8$ by Yu et al. \cite{yu_universal_2009}. In most other extraction methods of the bosonic mode energy, the normal state DOS is not respected, which is why, depending on the method, the $\Delta_0$ used there is most similar to what is here called $\Delta_\mathrm{max}$ or $\bar{\Delta}$. $\Delta_\mathrm{max}$ is the largest gap value that contributes to the elastic conductance spectrum and $\bar{\Delta}$ the momentum averaged and DOS weighted gap. 

\section{The case for $\mathrm{YBa_2Cu_3O_{6+x}}$}\label{sec:YBCO}
\begin{figure}[t]
    \centering
    \includegraphics{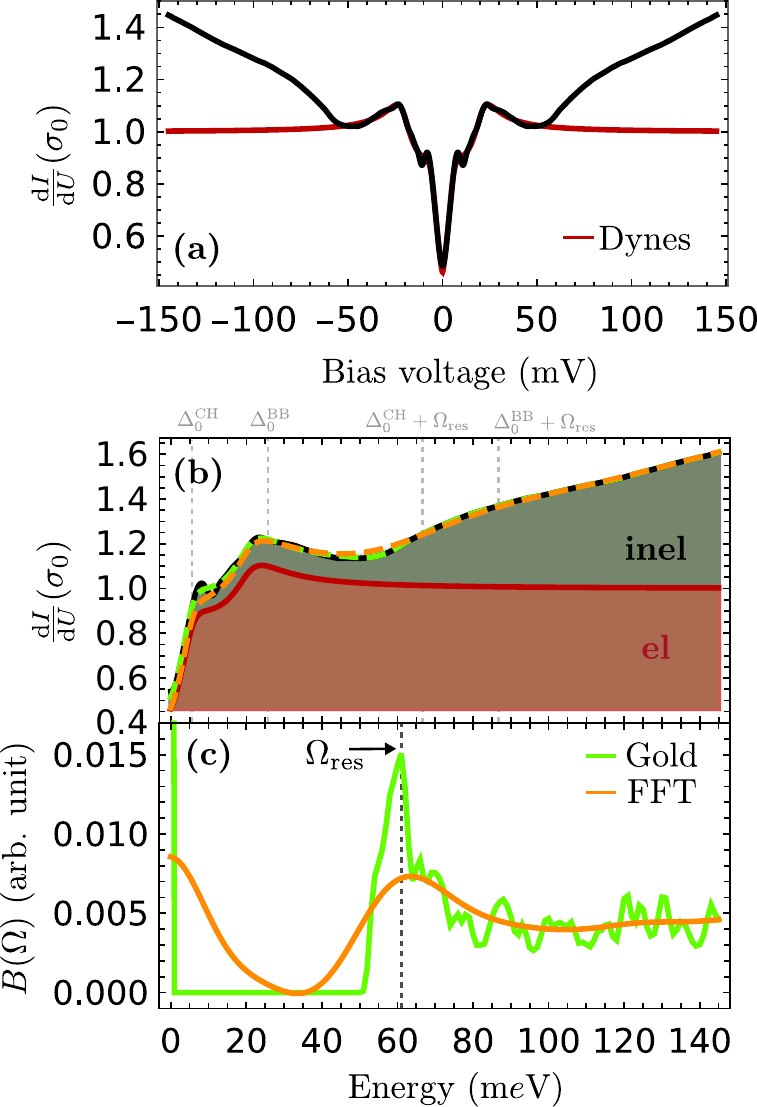}
    \caption{\textbf{Bosonic Spectrum Extraction for Y123}: (a) A generalized Dynes model (Eq.~(\ref{eq:sc_dos_el})) with $\Delta^\mathrm{AB}_0 = \SI{20.62}{\milli e\volt}$ for the anti-bonding (AB), $\Delta^\mathrm{BB}_0=\SI{25.77}{\milli e\volt}$ for the bonding (BB) and $\Delta^\mathrm{CHSS}_0=\SI{5.66}{\milli e\volt}$ for the chain (CH$_\mathrm{SS}$) band (red) was fitted to the experimental differential conductance (black). (b) The total conductance (black line) has been scaled up by the factor $\eta^{-1}=1.11$. The inelastic part of the conductance (vertical width of grey area) is given by the difference between the total (black) and the elastic part (red) of the conductance. Forward convolved conductance with the obtained boson spectral functions from the direct FFT method/Gold algorithm are shown in orange/green dashed lines. (c) Boson spectral function determined by direct FFT method/Gold algorithm (orange/green). The result of the direct FFT method has been regularized for clarity. A clear resonance mode at $\Omega_\mathrm{res}\approx \SI{61}{\milli e \volt}$ is visible. Zero-energy contributions are an artefact from the scaling procedure.}
    \label{fig:Fitting2}
\end{figure}
\subsection{Experimental results}
The d$I$/d$U$ spectrum for superconducting Y123 in Fig.~\ref{fig:Fitting2}(a) is qualitatively in excellent agreement with previous STM measurements \cite{maggio-aprile_direct_1995} and shows three low-energy features: (i) a superconducting coherence peak at $\approx \SI{25}{\milli e \volt}$ that is sharper than in Bi2212, (ii) a high-energy shoulder of the coherence peak and (iii) a low-energy peak at $\approx \SI{10}{\milli e\volt}$. The high-energy shoulder as well as the sub-gap peak are believed to arise from the proximity-induced superconductivity in BaO planes and CuO chains \cite{derro_nanoscale_2002, schabel_angle-resolved_1998, miller_density--states_1993, tachiki_tunneling_1990}. This would certainly account for the fact that these states are missing in the Bi-based compounds and that the sub-gap peak shows a direction-dependent dispersion in ARPES data \cite{damascelli_angle-resolved_2003, lu_superconducting_2001}. 
At energies larger than $\Delta$, we again find a clear dip-hump feature, similar as in Bi2212. The hump lies at $\approx \SI{60}{\milli e \volt}$. The V-shaped background conductance in the high-energy regime of the superconducting spectrum is in agreement with the predicted inelastic contribution by magnetic scattering in the spin-fermion model \cite{kirtley_inelastic-tunneling_1990, hlobil_tracing_2017}.

\subsection{Extraction of the bosonic spectrum}
We proceeded as in the case for Bi2212, but incorporated the one-dimensional band from the CuO chains as well as the bonding and anti-bonding band from the CuO$_2$ planes into the calculation of the normal state DOS to remodel the sub-gap peak and coherence peak shoulder in the estimated $\sigma^\mathrm{el}$ of the superconducting state. The optimal Dynes fit with gaps $\Delta^\mathrm{AB}_0 = \SI{20.62}{\milli e\volt}$ for the anti-bonding (AB), $\Delta^\mathrm{BB}_0=\SI{25.77}{\milli e\volt}$ for the bonding (BB) and $\Delta^\mathrm{CHSS}_0=\SI{5.66}{\milli e\volt}$ for the chain band is shown in red in Fig.~\ref{fig:Fitting2}(a). Because vacuum-cleaved surfaces favour tunneling into states of the CuO chain plane \cite{edwards_energy_1992, edwards_modulations_1994, derro_nanoscale_2002}, the sub-gap peak is pronounced and the contribution to the total DOS of the CH$_\mathrm{SS}$ band is, in our analysis, roughly five times higher than for the AB and BB band. 
The size of $\Delta^\mathrm{BB}$ is in good agreement with other scanning tunneling spectroscopy results spanning around 20 experiments, in which the extracted gap value lies between $\Delta = 18-\SI{30}{\milli e \volt}$ for optimally doped samples \cite{fischer_scanning_2007}. \black{For comparison: From Raman spectra, $\Delta_0$, i.e. the gap in the antinodal direction, is frequently found to be $\SI{34}{\milli e\volt}$ for optimally doped Y123 samples \cite{chen_oxygen-concentration_1993, gallais_evidence_2002, sugai_carrier-density-dependent_2003, le_tacon_investigations_2007}.} It should be noted that vacuum cleaved surfaces of Y123 tend to be overdoped \cite{edwards_energy_1992} which goes hand in hand with a steep decline of $\Delta$. The reason for discrepancy between the gap measured in STS and ARPES \cite{nakayama_bulk_2007, nakayama_doping_2009} \black{(also yielding $\Delta_0 \sim \SI{34}{\milli e\volt}$)} is expected due to two factors: (i) Although less influenced by a local gap variation than Bi2212, the gap of Y123 is expected to be inhomogeneous on a wider scale of $> \SI{100}{\nano\metre}$ \cite{nakayama_bulk_2007}. While ARPES yields an average gap over several of these domains, STS yields a more local gap. (ii) The measurement of a $k$-averaged gap value in STS naturally tends to give smaller values for a $d$-wave superconductor than the maximum gap size measured in ARPES. We try to eliminate this last effect by respecting the $k$-dependence of $\Delta$ and $\nu_\mathrm{n}^\mathrm{F}$ in our fit. Nevertheless, despite the large $T_c$, the spectroscopic results on Y123 in this work do not support an effective gap value of $>\SI{30}{\milli e\volt}$ because the total conductivity is already on the decrease at this energy.

Analogous to the case of Bi2212, the elastic part was, as a first guess, approximated by the Dynes fit to the total conductance times a scalar factor $\eta$. Here, $\eta = 0.9$ was chosen in order to secure the constraint $\sigma^\mathrm{inel}(e|U|>0)>0$. The (in)elastic parts to the total conductance are shown in red(grey) in Fig.~\ref{fig:Fitting2}(b).

We compare the extracted bosonic DOS obtained from direct deconvolution and Gold algorithm for Y123 in Fig.~\ref{fig:Fitting2}(c). The resonance mode at $\Omega_\mathrm{res}\approx \SI{61}{\milli e \volt}$ is significantly higher in energy than experimentally found by INS in (nearly) optimally doped samples with $\Omega_\mathrm{res}\approx \SI{41}{\milli e \volt}$ \cite{rossat-mignod_neutron_1991, bourges_spin_2000, hinkov_two-dimensional_2004, pailhes_resonant_2004} and even lies at the onset of the spin scattering continuum at $\Omega_c \approx \SI{60}{\milli e \volt}$ \cite{pailhes_resonant_2004}. Apart from the $k$-space integration, which shifts the peak centre to higher energies, several other factors can play a key role: (i) The well-studied $\SI{41}{\milli e\volt}$ odd-parity mode is paired with an even-parity mode at $\Omega^\mathrm{e}_\mathrm{res}\approx 53-\SI{55}{\milli e\volt}$ \cite{pailhes_doping_2006, pailhes_resonant_2004, pailhes_two_2003} which may be of the same origin as it vanishes at $T_c$. This mode appears with a $\approx 3-20$ times lower intensity in INS than the odd-parity mode, but this does not necessarily have to hold for a tunneling experiment. (ii) The bosonic spectrum extracted here is essentially poisoned by phononic contributions from every $k$-space angle. A disentanglement of phononic and electronic contributions to the total bosonic function by non-equilibrium optical spectroscopy showed that for $\Omega>\SI{100}{\milli e\volt}$ the bosonic function is purely electronic, yet in the energy range of the spin resonance, the contribution of strong-coupling phonons is almost equal to that of electronic origin \cite{dal_conte_disentangling_2012}. (iii) Apart from physical arguments, there can also be made sceptical remarks on the deconvolution procedure: Evidently, it heavily depends on the guess of the elastic tunneling conductance, which in this case does not contain strong-coupling features from an Eliashberg theory. (iv) The pronounced contribution of the CuO chains to the total conductance essentially causes the resonance mode to appear at roughly $\omega_\mathrm{hump}-\Delta^\mathrm{CHSS}$ instead of $\omega_\mathrm{hump}-\Delta^\mathrm{BB}$. Correcting for the \SI{20}{\milli e \volt} difference between the two gaps, it is likely that without sensitivity to the CuO chain gap, our extraction procedure will yield $\Omega_\mathrm{res}\approx \SI{41}{\milli e\volt}\approx 1.6\,\Delta^\mathrm{BB}_0 \approx 1.8\,\Delta^\mathrm{BB}_\mathrm{max}\approx 2.4\,\bar{\Delta}^\mathrm{BB}$.

\section{Conclusion}
We recorded scanning tunneling spectra on superconducting Bi2212 (UD82) and Y123 (OP92) at \SI{0.7}{\kelvin} and revealed a clear dip-hump structure outside the superconducting gap in both cases. The origin of this spectral feature can be traced back to a sharp resonance in the effective tunnel Eliashberg function. A careful separation of elastic and inelastic tunneling contributions enabled us to extract the bosonic excitation spectrum including this resonance.
%and an overall shape that shows clear resemblance to the expected generalized glue function due to spin scattering. %rework this
Comparing the obtained bosonic spectrum with inelastic neutron scattering data yields good agreement with the observed resonance mode and supports that magnetic fluctuations play an important role in the pairing mechanism of the cuprate superconductors.

%That this is possible for a nodal superconductor despite the complications of a highly ill-posed deconvolution problem and loss of explicit momentum information showcases the dominance of the Fermi surface hot spots in tunneling spectra. %rework this

Our extraction method of the bosonic spectrum from scanning tunneling spectra paves a way to complement glue functions determined from optical spectroscopy or ARPES and has several advantageous features: The usage of scanning tunneling spectra yields the option for atomic resolution of the bosonic modes on the superconductor surface \cite{lee_interplay_2006, chi_imaging_2017, jandke_unconventional_2019} as well as easy access of both occupied and unoccupied quasiparticle states with the high energy resolution of cryogenic STM setups.

\section*{Acknowledgements}
The authors acknowledge funding by the Deutsche Forschungsgemeinschaft (DFG) through CRC~TRR~288~-~422213477 ``ElastoQMat'', project A07, B03 and B06.
%The authors acknowledge funding by the Deutsche Forschungsgemeinschaft (DFG) under the grant WU 394/12-1 and SCHM 1031/7-1.

% \section{Author Contributions}
% T.W. grew the Y123 and Bi2212 single crystals. T.G. and M.H. conducted the STS measurements. T.G. analysed the data, performed model calculations and the bosonic spectrum extraction by deconvolution. T.G. wrote the manuscript including input from all authors.  Discussions with M.L.T. and J.S. helped to navigate the scientific field both experimentally and theoretically and drastically improved the manuscript. W.W. headed this study and J.S. laid the theoretical groundwork. 
\appendix
\section*{Appendix}
\renewcommand{\thesubsection}{\thesection\Alph{subsection}}
\subsection{Calculated normal density of states}\label{sec:app/calculated normal dos}
\begin{table}[b]
    \centering
    \caption{\textbf{Tight binding parameters}: Chemical potential $\mu$ and hopping parameters $t_i$ used in the dispersion relation for the Bi2212 and Y123 bands (Eqs.~\eqref{eq:tight-binding dispersion} and \eqref{eq:tight-binding dispersion 1D chain}).}
    \label{tab:tb parameters}
    \begin{tabular}{c|c|c|c|c|c|c|c}
        Band& $\mu$ & \multicolumn{2}{c|}{$t_1$} & $t_2$ & $t_3$ & $t_4$ & $t_5$ \\[1pt]\hline\hline  \rule{0pt}{1\normalbaselineskip} 
        &\multicolumn{7}{c}{Bi2212}\\[1pt]\hline\hline  \rule{0pt}{1\normalbaselineskip}  
        BB & -0.1305 & \multicolumn{2}{c|}{-0.5951} & 0.1636 & -0.0519 & -0.1117 & 0.051 \\[1pt]\hline \hline\rule{0pt}{1\normalbaselineskip} 
        &\multicolumn{7}{c}{Y123}\\\hline\hline\rule{0pt}{1\normalbaselineskip}
        BB & -0.38 & \multicolumn{2}{c|}{-1.1259} & 0.5540 & -0.1774 & 0.0701 & 0.1286 \\
        AB & -0.515 & \multicolumn{2}{c|}{-1.0939} & 0.5112 & -0.0776 & -0.1041 & 0.0674 \\\hline \rule{0pt}{1\normalbaselineskip}    &&$t_a$&$t_b$& \multicolumn{4}{c}{}\\\hline\rule{0pt}{1\normalbaselineskip}
        CH$_\mathrm{SS}$ & -0.2155 & -0.12 & -0.0035 &\multicolumn{4}{c}{-}   
    \end{tabular}
\end{table}
\begin{figure}[t]
    \centering
    \includegraphics[scale=1]{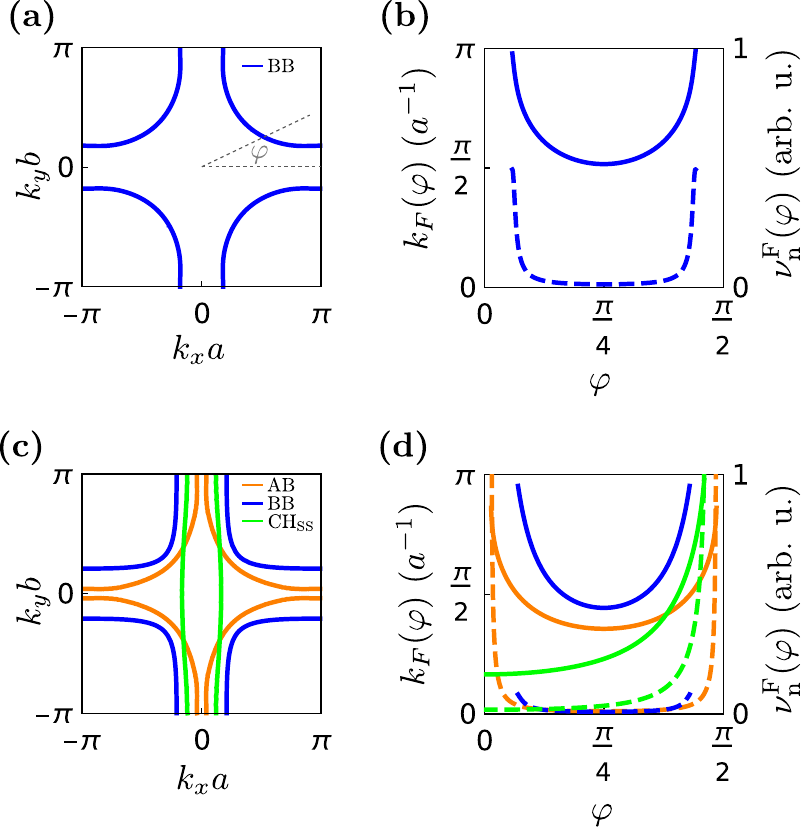}
    \caption{\textbf{Normal State Electrons}: (a,c) Calculated Fermi surface in the first 2D BZ of Bi2212 (a) and Y123 (c). (b,d) Calculated Fermi wave vector $k_F(\varphi)$ (solid line) and normal conducting DOS along the Fermi surface contour $\nu_\mathrm{n}^\mathrm{F}(\varphi)$ (dashed line) as function of polar angle in the first BZ quadrant (sketched in a) for Bi2212 (b) and Y123 (d). Colors match the Fermi surface contours of the individual bands (AB, BB, CH$_\mathrm{SS}$) from (a,c).}
    \label{fig:NormalStateElectrons}
\end{figure}
For Bi2212 and Y123, the in-plane dispersion of the CuO$_2$ planes is described by a tight binding model of the form 
\begin{align}
    \epsilon(k_x,k_y) &= \frac{t_1}{2}(\cos(k_x)+\cos(k_y)) + t_2\cos(k_x)\cos(k_y) \nonumber\\&
    +\frac{t_3}{2}(\cos(2k_x)+\cos(2k_y))\nonumber +\frac{t_4}{2}(\cos(2k_x)\cos(k_y)\nonumber\\
    &+\cos(k_x)\cos(2k_y))+t_5\cos(2k_x)\cos(2k_y) \nonumber\\
    &-\mu \label{eq:tight-binding dispersion}
\end{align}
with chemical potential $\mu$ and hopping parameters $t_i$ as proposed in Ref.~\cite{norman_phenomenological_1995}. For Bi2212, we used the set of parameters from Ref.~\cite{norman_phenomenological_1995} for a near optimally doped crystal and for Y123 we started from the parameters proposed in Ref.~\cite{schabel_angle-resolved_1998} for the optimally doped case and adjusted chemical potential, as well as $t_2$ to fit recently obtained Fermi surface contours measured by ARPES \cite{iwasawa_surface_2018}. While we only consider the binding band (BB) for Bi2212, for Y123, we take the binding (BB), anti-binding (AB) and the chain band (CH$_\mathrm{SS}$) into consideration. The latter is modeled by a dispersion of the form 
\begin{align}
    \epsilon(k_y,k_y) = 2t_a\cos(k_x)+2t_b\cos(k_y)-\mu.\label{eq:tight-binding dispersion 1D chain}
\end{align}
The used tight binding parameters are summarized in Tab.~\ref{tab:tb parameters}. The calculated Fermi surface in the first Brillouin zone (BZ) is shown in Fig.~\ref{fig:NormalStateElectrons}(a,c) for Bi2212 and Y123. An analytic expression for the Fermi wave vector $k_F(\varphi)$ is retrieved from the solution of $\epsilon(k,\varphi)=0$ where $\epsilon(k,\varphi)$ is the polar representation of Eq.~(\ref{eq:tight-binding dispersion}). The normal DOS is then given by
\begin{equation}
    \nu^\mathrm{F}_\mathrm{n} = \oint \frac{\mathrm{d}^2k}{|\bm{\nabla}_{\bm{k}} \epsilon|} \rightarrow
    \int_l \frac{\mathrm{d}\varphi}{\bm{\nabla}_{k,\varphi}\epsilon(l(\varphi))}\norm{l^\prime(\varphi)}\mathcomma
    \quad
    \varphi \in M_\varphi
\end{equation}
where $l(\varphi) = (k_F(\varphi)\cos(\varphi), k_F(\varphi)\sin(\varphi))^T$ is a parametrization of the path along the Fermi surface, $\norm{\cdot}$ is the euclidean norm and $M_\varphi = \{\varphi|k_F(\varphi)\in \mathrm{1.BZ}\}$. It is shown as a function of the polar angle $\varphi$ in Fig.~\ref{fig:NormalStateElectrons}(b,d) for Bi2212 and Y123 respectively.

\subsection{The numerical scaling factor $\eta$}\label{subsec: app/The numerical scaling factor}
\begin{figure}
    \centering
    \vspace{1.5mm}
    \includegraphics[scale=0.8]{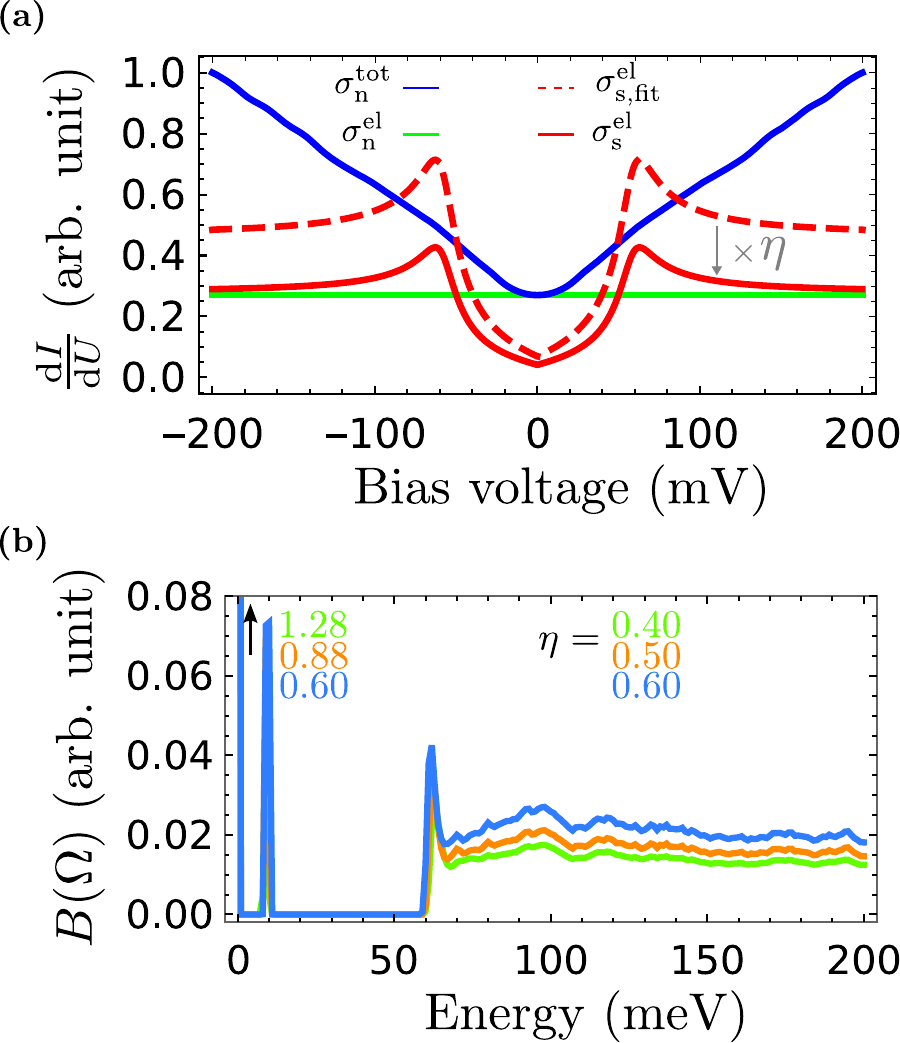}
    \caption{\textbf{Numerical Scaling Factor}: (a) Determination of $\eta$ through the boundary condition
$\int_{-\infty}^\infty \mathrm{d}\omega \sigma_\mathrm{n}^\mathrm{el}=\int_{-\infty}^\infty \mathrm{d}\omega \sigma_\mathrm{s}^\mathrm{el}$ . (b) Variation of $\eta$ leaves the general shape of the extracted bosonic spectrum unaffected except for the magnitude of its zero-energy peak.}
    \label{fig:NumericalScalingFactor}
\end{figure}
We used the boundary condition 
$$\int_{-\infty}^\infty \mathrm{d}\omega\, \sigma_\mathrm{n}^\mathrm{el}=\int_{-\infty}^\infty \mathrm{d}\omega\, \sigma_\mathrm{s}^\mathrm{el}$$
to approximate $\eta$, i.e. the total number of electronic states is conserved in the phase transition from the normal to the superconducting phase. This procedure is depicted in Fig.~\ref{fig:NumericalScalingFactor}(a).

In order to make sure that the introduction of the numerical scaling factor $\eta$ has no poisoning effect on our extracted bosonic spectrum, the deconvolution of the Bi2212 spectrum by Gold's algorithm was performed for four different values of $\eta$. The results shown in Fig.~\ref{fig:NumericalScalingFactor}(b) are comforting in the sense that the overall shape of the bosonic spectrum is unchanged. The only major difference lies in the magnitude of the zero-energy peak which is to be expected from a scalar multiplication, but since this peak is anyhow out of the bounds of physical contributions it does not harm the analysis.

\subsection{LDOS inhomogeneity}\label{subsec: app/ldos inhomo}
\begin{figure}[t]
    \centering
    \vspace{1.5mm}
    \includegraphics[scale=1]{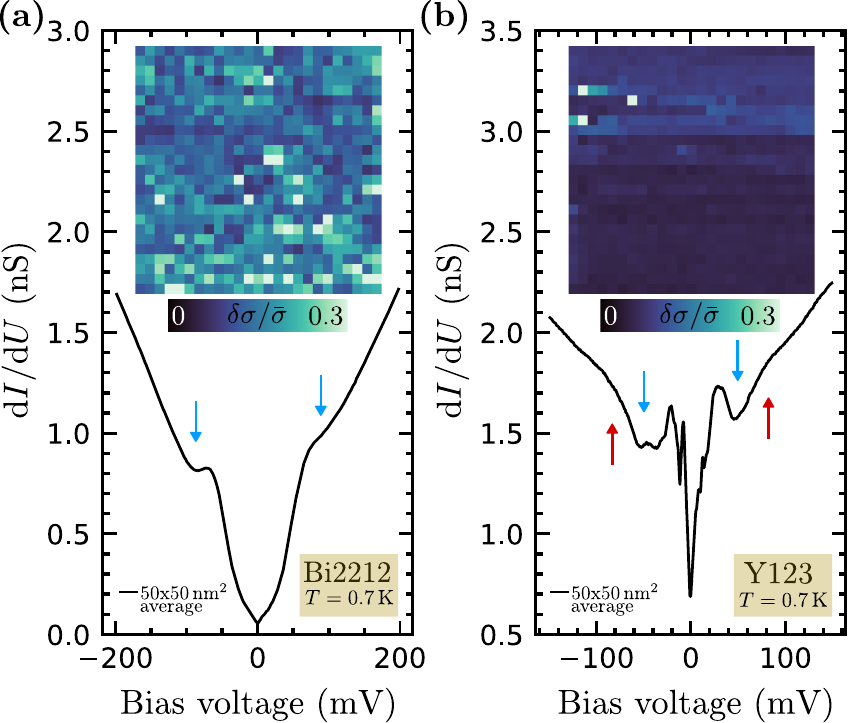}
    \caption[Conductance Inhomogeneity on Superconducting Bi2212 and Y123]{\textbf{LDOS Inhomogeneity}: Position averaged bias spectra on a $50\times \SI{50}{\nano\metre}^2$ area at $T=\SI{0.7}{\kelvin}$ for Bi2212 (a) and Y123 (b). Heat maps in the inset show the variation of the differential conductance within the averaging area. The higher inhomogeneity of the Bi2212 surface is reflected in both the conductance variation map and the blurred position averaged spectrum. The characteristic dip and hump are marked by blue and red arrows.}
    \label{fig:Inhomogeneity}
\end{figure}
As reported by Fischer et al., Bi2212 tends to show a large inhomegeneity of its LDOS in the superconducting state \cite{fischer_scanning_2007}. This can be confirmed in our experiment by direct comparison of the conductance inhomegeneity measured on Bi2212 and Y123, shown in Fig.~\ref{fig:Inhomogeneity}. The heat maps of the conductance variation 
\begin{equation}
    \delta \sigma / \bar{\sigma}(x,y) =\frac{1}{N} \sum_{U=-U_t}^{U_t} \frac{\sigma(eU,x,y)-\bar{\sigma}(eU)}{\bar{\sigma}(eU)}
\end{equation}
show that it is about three times higher on the Bi2212 surface than on the Y123 surface. As a consequence, a position averaged spectrum over a $50\times\SI{50}{\nano\metre}^2$ area can preserve detailed gap features better for Y123 than for Bi2212. Especially the dip-hump (dip marked by blue, hump marked by red arrows in Fig.~\ref{fig:Inhomogeneity}) feature is still clearly visible in the position averaged spectrum of Y123 at $\epsilon\approx \SI{60}{\milli e \volt}$ but is invisible in Bi2212. The preservation of this feature in the spectrum is crucial for our ITS analysis. Therefore, in the case of Bi2212, an average spectrum at one specific location, at which the dip-hump spectral feature was clearly visible, was chosen for this study. For Y123, the position averaged spectrum was chosen. 

\newpage

\bibliography{main.bib}

%\noindent
%{\bf Author contribution}

%\noindent
%{\bf Author information}
% Correspondence and requests for materials should be addressed to T.G. and W.W. (wulf.wulfhekel@kit.edu and thomas.gozlinski@kit.edu).

\end{document}